# Magnetic field dependence of superconducting energy gaps in $YNi_2B_2C$: Evidence of multiband superconductivity

S Mukhopadhyay, Goutam Sheet[a], and P Raychaudhuri[b]

Department of Condensed Matter Physics and Materials Science, Tata Institute of Fundamental Research, Homi Bhabha Rd., Colaba, Mumbai 400005, India.

and

H Takeya

National Institute for Materials Science, 3-13 Sakura, Tsukuba, Ibaraki 305-0003, Japan

We present results of in field directional point contact spectroscopy (DPCS) study in the quaternary borocarbide superconductor $YNi_2B_2C$, which is characterized by a highly anisotropic superconducting gap function. For I||a, the superconducting energy gap ($\Delta$), decreases linearly with magnetic field and vanishes around 3.25T which is well below the upper critical field ($H_{c2}$~6T) measured at the same temperature (2K). For I||c, on the other hand, $\Delta$ decreases weakly with magnetic field but the broadening parameter ($\Gamma$) increases rapidly with magnetic field with the absence of any resolvable feature above 3.5T. From an analysis of the field variation of energy gaps and the zero bias density of states we show that the unconventional gap function observed in this material could originate from multiband superconductivity.

74.70.-b, 74.50.+r, 74.25.Ha, 74.70.Dd

---

[a] Corresponding Author: goutam@tifr.res.in

[b] Corresponding Author: pratap@tifr.res.in

The quaternary borocarbide superconductor $YNi_2B_2C$ has received renewed interest in the last couple of years due to the unusual anisotropy in the gap function observed in this material. A variety of recent experiments[1,2,3,4,5,6,7,8] give strong evidence of sharp minima in the gap function along certain *k*-directions with possible point nodes along [100] and [001]. Based on the shape of the gap function an order parameter symmetry with mixed angular momentum, namely, *s+g* symmetry[9], has been proposed for this compound. Though *s+g* symmetry seems to be consistent with experimental observations of the gap anisotropy at low temperatures, the origin of this symmetry still remains debatable. Moreover, in the *s+g* descriptions so far proposed, a roughly spherical FS is implicitly assumed where the multiband nature of the complex Fermi surface[10,11,12,13] (FS) in $YNi_2B_2C$ remains largely ignored.

An alternative scenario has also been explored in $YNi_2B_2C$ where the gap anisotropy could originate from different bands on the Fermi surface having different gap values due to difference in their coupling strengths. Band structure calculations as well as de Haas-van Alphen studies reveal that $YNi_2B_2C$ has a multiply connected FS extending over 3 bands[13]. In measurements, such as thermal and electrical conductivity, tunneling, optical spectroscopy and point contact spectroscopy, the contribution from a particular band in a particular *k*-direction depends on a weighted average of the Fermi velocity ($v_{Fk}$) and the density of states ($N_k(E_F)$). Therefore for a non-spherical Fermi surface the contribution of different bands could be different in different *k* directions. Recent point contact spectroscopy studies have shown that such a situation indeed exists in the two band superconductor $MgB_2$, where the relative contribution of the π and σ bands in the point contact current is different when the current is injected along the *a* and *c* directions respectively[14]. In YNi$_2$B$_2$C, from an analysis of the temperature dependence of H$_{c2}$, it has been proposed[15] that both strongly coupled slow electrons as well as weakly coupled fast electrons co-exist on the FS. The existence of weakly coupled Fermi surface pockets with small or zero gap is further corroborated from quantum oscillation measurements, where one of the oscillation frequencies arising from one particular extremal orbit persists deep in the superconducting state[16] whereas the others damp out below 0.8H$_{c2}$. Therefore the possible role of the multiband nature of the FS on the observed gap anisotropy in this material needs to be explored more carefully.

DPCS, namely where the point contact spectrum is recorded by injecting current along different crystal directions is a powerful tool to investigate the gap anisotropy in unconventional superconductors. In contrast to measurements such as the angular variation of thermal conductivity or specific heat in applied magnetic field, this technique allows a direct measurement of the superconducting energy gap in different *k*-direction over a wide temperature and magnetic field range. In this paper we report a DPCS study of the magnetic field dependence of the superconducting energy gap along the two principal crystallographic axes ([001] and [100]) of a single crystal of YNi$_2$B$_2$C, by injecting current (I) either along *a* or *c*, up to a field of 7T. Our key observations are the following: (i) For I||*a*, the small gap ($\Delta_{I||a}$) decreases rapidly with magnetic field (H) and vanishes at H~3.25T, which is much smaller than the H$_{c2}$ of the superconductor; (ii) the large gap ($\Delta_{I||c}$) for I||*c*, on the other hand, decreases much more slowly with magnetic field. The magnetic field dependence of the two gaps as well as the zero bias density of states extracted from directional point contact spectra are in good agreement with the predictions of a two-band superconductor with weak interband impurity scattering.

High quality singly crystals of $YNi_2B_2C$ grown by traveling solvent floating zone method were from the same batch as the one used[17] in ref. 3. The measurements were carried out on relatively large crystals (0.5mm×0.5mm×2mm) with well-defined facets along [100] and [001]. For point contact measurements a mechanically cut fine silver tip was brought in contact with the [100] or [001] facet (for I||$a$ and I||$c$ respectively) of the crystal using a 100 threads per inch differential screw arrangement in a liquid He cryostat. The differential conductance (dI/dV vs. V) of the point contact was measured directly using a four-probe current modulation technique. For all the spectra reported in this paper the point contact resistance was in the range 10-20Ω ensuring that the contacts were in the ballistic limit[18]. For both current directions, the magnetic field (H) was applied parallel to the current. Since the upper critical field ($H_{c2}$) in $YNi_2B_2C$ is slightly different for H||$a$ and H||$c$, the upper critical fields for both field directions were determined from the field variation of ac susceptibility ($\chi$ vs. H) measured at 15kHz in a home made ac susceptibility setup.

Figure 1(a)-(b) show typical point contact spectra for I||$a$ and I||$c$ at 2.3K measured at different applied magnetic fields. The spectra were fitted with the Blonder-Tinkham-Klapwijk (BTK)[19] model using the superconducting energy gap Δ, the barrier height coefficient Z, and a broadening parameter[20] Γ as fitting parameters.

Γ phenomenologically accounts for the broadening in the superconducting density of states (DOS) from its ideal BCS value. This broadening could arise either from a finite lifetime of the superconducting quasiparticles arising from impurity scattering[20] or from a distribution of superconducting gap values[21]. The broadened DOS is given in terms of Δ and Γ as,

$$N(E) \sim \text{Re}\left(\frac{E+i\Gamma}{\sqrt{(E+i\Gamma)^2 - \Delta^2}}\right) \quad (1)$$

In zero field, the best-fit parameters for I||c and I||a are $\Delta_{I||c}$~2.2meV, Z=0.585, $\Gamma_{I||c}$=0.2, and $\Delta_{I||a}$~0.37meV, Z=0.63, $\Gamma_{I||a}$=0.145 respectively. The large value of $\Delta_{I||c}$ compared to $\Delta_{I||a}$ is consistent with our earlier observations[3], though the gap anisotropy is somewhat larger ($\Delta_{I||c}/\Delta_{I||a}$~6) than previously reported. This difference is possibly due to our inability to precisely control the direction of the current due to surface roughness. Consistent with our earlier observation the relative broadening (Δ/Γ) in zero field is also different for the two directions[22], ($\Gamma_{I||c}/\Delta_{I||c}$)~0.09 and ($\Gamma_{I||a}/\Delta_{I||a}$)~0.392.

Before discussing the significance of these observations we briefly summarize the mechanism of ballistic transport in a multiband metal. When a current (*I*) is injected through a ballistic interface between two metals, the net flux of electrons from a particular band *i* on the FS is given by[23],

$$I_i \propto \oint_{FS} N_{ik}(\bar{v}_{ik} \cdot \hat{n}) dS_F = \langle N_{ik} v_{ik\hat{n}} \rangle_{FS} \quad (2),$$

where ***k*** is an wave-vector at the FS, $N_{ik}$ is the density of states, $\bar{v}_{ik} \cdot \hat{n}$ $(= v_{ik\hat{n}})$ is the component of the Fermi velocity along $\hat{n}$ and $dS_F$ is an elementary area on the FS[24]. The quantity within $\langle ... \rangle$ is easily seen to be $S_{i\hat{n}}$, the area of projection of the *i*-th band on the interface plane. The total current (*I*) is therefore given by a sum over all bands namely, $I = \sum_i I_i$, where the contribution from each band is proportional to its area of projection on the interface plane. In general for a non-spherical FS, the area of projection of a given band will be different along different directions. If the FS extends over several Brillouin zones, the contribution of different Fermi surface sheets to the current will be different in different directions. In a point contact Andreev reflection experiment where the observed spectrum will contain contribution from different bands in proportion to their projection area on a plane perpendicular to the current flow, the average value of $\Delta$ when $I \parallel \hat{n}$ will be given by a weighted average of the form $\dfrac{\sum_i \langle \Delta_{ik} N_{ik} v_{i\hat{n}k} \rangle_{FS}}{\sum_i \langle N_{ik} v_{i\hat{n}k} \rangle_{FS}}$. Thus for an anisotropic FS, if the superconducting energy gap is different in different bands, different values of $\Delta$ in different directions is expected to arise from unequal contribution of different bands[25]. In $YNi_2B_2C$, where the two dimensional nature of the atomic structure gives rise to spheroidal, square and cylindrical FS[12], a significant difference in the relative contribution of different bands could be expected when measured along *c* and *a*.

We now focus our attention on the zero field data at low temperatures. The large anisotropy between $\Delta_{I \parallel a}$ and $\Delta_{I \parallel c}$ could arise from both scenario discussed before, namely, a superconductor with mixed angular momentum ($s+g$) symmetry or multiband superconductivity. In the former case the observed anisotropy would arise due to the gap being anisotropic over a roughly spherical FS due to anisotropic pairing interactions. In the second case the observed anisotropy would arise from a FS sheet with large gap having predominant contribution in the point contact current for I||$c$, whereas for I||$a$ a sheet with small gap has a bigger contribution. Therefore based on the extracted parameters in zero field data alone, it is not possible to distinguish between a mixed angular momentum ($s+g$) superconductor and multiband superconductivity even when the values of $\Gamma_{I||a}$ and $\Gamma_{I||c}$ are taken into consideration. In zero fields the most likely origin of the broadening parameter is a distribution of the superconducting energy gaps seen by the point contact current. This could again arise from both origins, namely, from different energy gaps on different bands of the anisotropic multiband FS or from the distribution of energy gaps over an isotropic FS due to unconventional pairing symmetry.

To gain further insight on the origin of the aniostropic gap, we now look at the field dependence of the point contact spectra. To find out the effect of magnetic field we fit the in-field spectra for both current directions, using the same equations as in zero field, keeping the value of Z constant to the corresponding zero field values and varying the values of $\Delta$ and $\Gamma$. The field variation of $\Delta_{I||c}$ and $\Delta_{I||a}$ as well as the corresponding broadening parameters ($\Gamma_{I||c}$ and $\Gamma_{I||a}$) are shown in figure 2(a) and (b). $\Delta_{I||a}$ decreases rapidly with magnetic field to about 30% its zero field value at 2.25T. Beyond this field value we cannot conclude anything regarding the gap in this direction. However, a linear extrapolation of the data shows that $\Delta_{I||a}$ should vanish at about 3.25T which is much lower than the corresponding $H_{c2}(H\|a)$~6T [26]. $\Delta_{I\|c}$, on the other hand shows a slow decrease with magnetic field[27] and reaches about 70% of its zero field value at 3.25T ($H_{c2}(H\|c)$~5.25T)[26]. The broadening parameter $\Gamma_{I\|c}$, however, increases rapidly with magnetic field and beyond 3.25T, the point contact spectra for $I\|c$ becomes too broad to resolve the superconducting energy gaps. The variation of $\Delta_{I\|a}$ and $\Delta_{I\|c}$ are very similar to the small and large gap in the two-band superconductor[28] $MgB_2$ where the small gap vanishes at about 1T whereas the large gap is almost constant in this field range. Furthermore, $\Delta_{I\|c}/k_BT_c$~3.6 is close to the weak coupling BCS value ($\Delta/k_BT_c$~3.52) suggesting that $T_c$ (and $H_{c2}$) in $YNi_2B_2C$ is governed by the large gap. This again is similar to $MgB_2$ and is expected for a two-band superconductor with weak interband scattering.

Since at present we do not have a theoretical model which accounts for the multiband nature of superconductivity $YNi_2B_2C$, we qualitatively compare our data with the predictions of the vortex state of a two-band superconductor[29] with weak interband scattering[30,31] proposed in the context of $MgB_2$. The faster decrease of the small gap ($\Delta_{I\|a}$) with magnetic field compared to large one ($\Delta_{I\|c}$) is consistent with the prediction of a two band model[30] where the diffusion constant (defined as $D_i = 2\pi T_c \xi_i^2$ where $\xi_i$ is the coherence length of the $i^{th}$ band) of the band with small gap is 20% of that of the large gap. For the same parameters it was theoretically predicted that the zero energy DOS ($N(0)$) for the small gap will reach its normal state value for fields much smaller than $H_{c2}$. To compare this prediction with our experimental results we calculate the zero energy

DOS as a function of magnetic field from equation (1). The field variation of $N(0)$ extracted for the two current directions is shown in the inset of figure 3. For I∥$a$ a significant zero bias DOS, presumably arising from gapless regions of the FS is seen even for zero field. To look at the field variation we separate $N(0)$ into an intrinsic part ($N_i(0)$) and a field dependent part $N_H(0)=N(0)-N_i(0)$, where $N_i(0)$ is the zero energy DOS in zero field. Figure 3 shows the variation of $N_H(0)$ for I∥$a$ and I∥$c$ normalized to their respective values at the highest field, as a function of the reduced magnetic field, $h(=H/H_{c2})$. The striking similarity of the experimental data with the two-band prediction strongly suggests that the observed spectra for I∥$a$ and I∥$c$ originate from two different bands.

Finally, we focus our attention on the temperature dependence[32] of $\Delta_{I\|a}$ and $\Delta_{I\|c}$ (Fig. 4). While $\Delta_{I\|c}$, shows a nearly BCS temperature variation vanishing only at $T_c$~14.5K, $\Delta_{I\|a}$ decreases rapidly at low temperatures and goes below our measurement resolution above 6K. This behavior cannot be understood from a gap function of the form $\Delta(\bar{k})=\Delta_0 f(\bar{k})$, suggested for the $s+g$ scenario, where the temperature dependence of $\Delta(k)$ originates from the variation of $\Delta_0$ alone. In such a situation the temperature variation of $\Delta(k)$ along different $k$ direction should vary just by a multiplicative factor, unless the amplitudes of $s$ and $g$ components in the order parameter symmetry is temperature dependent. Such a temperature dependent amplitude however cannot be rationalized within the BCS theory[33]. On the other hand, for a weakly interacting two-band scenario the temperature dependence of the two bands will be different if the coupling strengths on the two bands are different[34]. To highlight this point we plot (solid lines) in figure 4 the expected variation for $\Delta_{I\|c}$ and $\Delta_{I\|a}$, assuming that the two gaps originate from two weakly coupled bands with no interband scattering, and have $T_c$ of 14.5K and 4.6K respectively. The dashed line in the same figure shows the expected variation of $\Delta_{I\|a}$ for a gap symmetry of form $\Delta(\bar{k}) = \Delta_0 f(\bar{k})$. At low temperatures, $\Delta_{I\|a}$ follows the solid line and deviates towards the large gap at higher temperatures. This is the behavior expected for a two band superconductor[34], with small but finite interband scattering.

We can now try to speculate on the FS sheets that are likely to be responsible for the large and small energy gaps. Based on band structure calculations[12], for non-magnetic borocarbides (Y/Lu)Ni$_2$B$_2$C, the 3 bands crossing the FS produce 5 Fermi surface sheets: Two ellipsoids centered at the Γ points, with their long axes parallel to *c*-axis, two square FS centered at P with sides parallel to [100] and [010], and a cylindrical FS constricted along [100] and [010] planes near to $k_z=\pi/c$. The broad structures of these calculations have been verified from different experiments, such as dHvA oscillations[13] and two-dimensional angular correlation of electron–positron annihilation radiation (2D-ACAR) technique[35]. In dHvA measurements, the frequency arising from the extremal orbit enclosing the ellipsoid sheet persists down to 3T suggesting that this sheet possibly remains gapless[16]. However, this Fermi sheet encloses only ~0.3% of the first Brillouin zone and plays a relatively minor role in point contact experiments. On the remaining two kinds of FS sheets, band structure calculations reveal that the Fermi velocity ($v_F$) varies[36] by a factor of 6. From an analysis of the temperature variation of $H_{c2}$, it has been suggested[15] that the slow electrons are strongly coupled and contribute primarily to superconductivity. Since, for the cylindrical FS the Fermi velocity is predominantly in the $k_x$-$k_y$ plane ($v_{F_z} \approx 0$), it can be seen from equation (2) that this band will primarily contribute in the current for I||*a*. (The contribution of this band for I||*c* will arise only from the deviation from its ideal cylindrical nature.) This surface primarily comprises of fast electron with only small pockets of slow electrons and is likely to contribute to the small gap observed for I||*a*. For I||*c*, on the other hand the predominant contribution is likely to arise from square FS centered at P, for which on most points on the Fermi surface, $\bar{v}_F$ has a large component along $\hat{z}$. On this FS, regions in the vicinity of $k_{Fz}$ primarily consist of slow electrons[36] which are likely to be responsible for the large gap observed[37] for I||*c*.

In summary, we have carried out a detailed study of the magnetic field and temperature dependence of the anisotropic superconducting energy gap in $YNi_2B_2C$ using directional point contact spectroscopy. Our results cannot be understood from a simple mixed angular momentum symmetry scenario, such as *s+g* previously suggested for this material. However, field dependence of Δ as well *N(0)* are in good agreement with theoretical predictions for a two-band superconductor, suggesting that the unusual gap anisotropy possibly originates from a multiband scenario, where different bands on the FS have different coupling strengths. The difference in the temperature dependence of the superconducting energy gap in different directions further supports the multiband scenario, suggesting that the role of the complex FS on the gap anisotropy in $YNi_2B_2C$ needs to be theoretically explored in detail.

*Acknowledgements:* We would like to thank Professors S Ramakrishnan and R Nagarajan for enlightening discussions, Professor A K Grover for encouragement and support and S P Pai for technical help. Two of us (SM and GS) would like to thank TIFR Endowment Fund for partial financial support.

Figure 1: Point contact spectra at 2.3K in different magnetic fields for (a) I||c and (b) I||a. The solid circles are experimental data and the solid lines are the BTK fits to the data. The magnetic fields for (a) are 0, 0.25, 0.75, 1.25, 1.75, 2.25, 2.75, 3.25T and for (b) are 0, 0.25, 0.375, 0.5, 0.75, 0.875, 1.0, 1.25, 1.5, 1.75, 2.25T. The conductance curves are normalized to their respective values at high bias.

Figure 2: Magnetic field dependence of (a) $\Delta_{I||c}$ and $\Delta_{I||a}$ and (b) $\Gamma_{I||c}$ and $\Gamma_{I||a}$ at 2.3K. The solid lines are guides to the eye.

Figure 3: The variation of the zero bias density of states ($N_H(0)$) with magnetic field after subtracting the zero field contribution for I||a and I||c as a function of the reduced magnetic field (h=H/H$_{c2}$). Inset: The total zero bias density of states ($N(0)$) as a function of magnetic field extracted from the point contact spectra for I||a and I||c.

Figure 4: Temperature variation of $\Delta_{I||c}$ (solid square) and $\Delta_{I||a}$ (solid circle) in zero magnetic field. The solid lines are the expected variation for the large and small gaps assuming their T$_c$ to be 14.5K and 4.6K respectively. The dashed line is the expected variation of the small gap if the gap anisotropy originates from a gap function of the form $\Delta = \Delta_0 f(\bar{k})$.

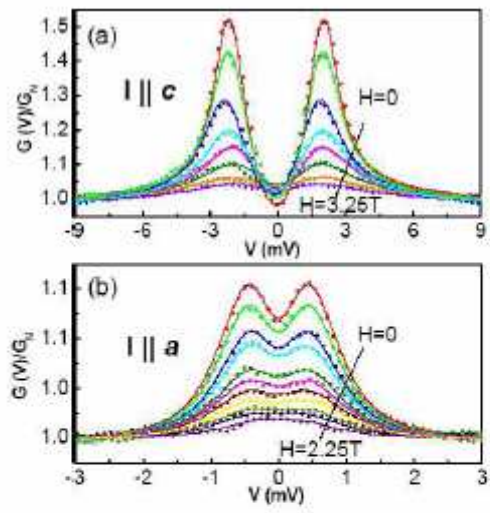

Figure 1
(Mukhopadhyay et al.)

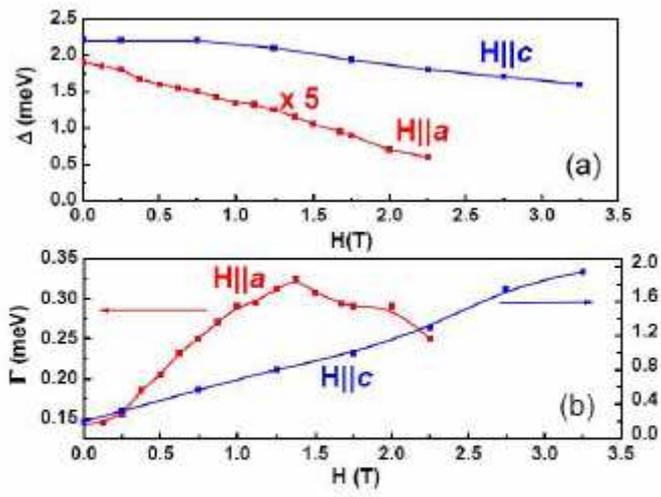

Figure 2
(Mukhopadhyay et al.)

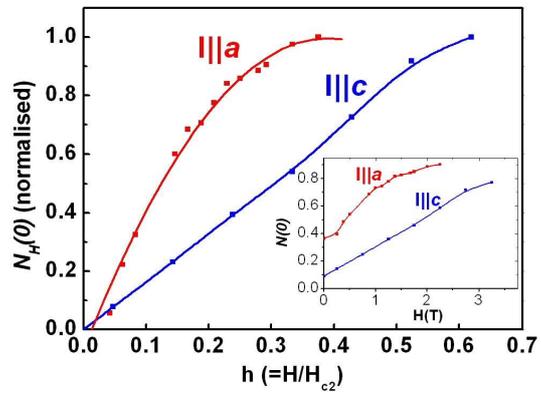

Figure 3
(Mukhopadhyay et al.)

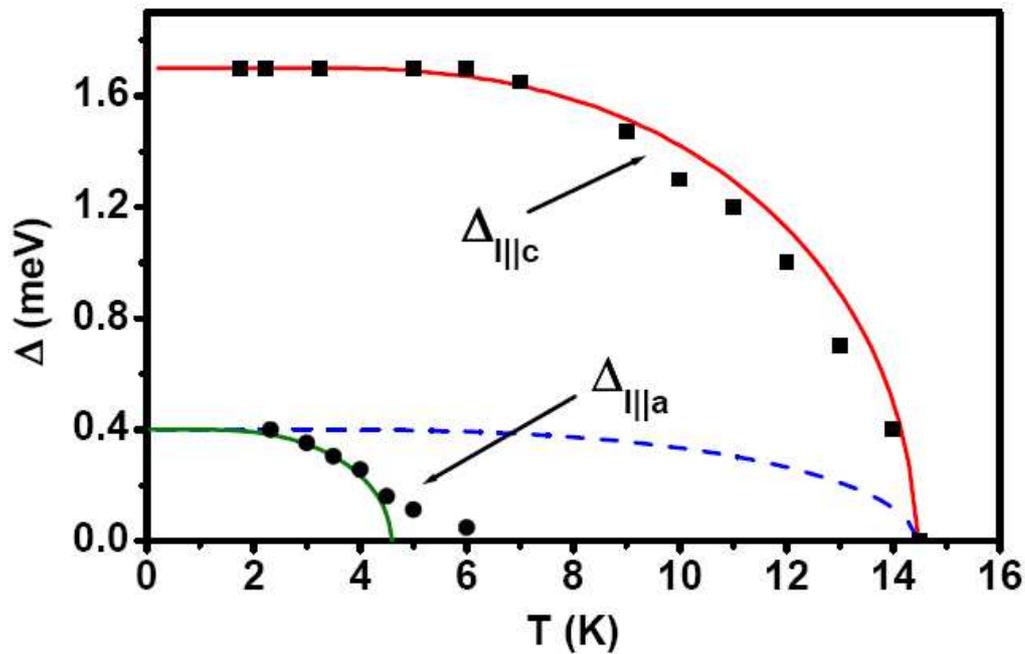

Figure 4 (Mukhopadhyay et al.)

[1] K Izawa, K Kamata, Y Nakajima, Y Matsuda, T Watanabe, M Nohara, H Takagi, P Thalmeier, and K Maki, Phys. Rev. Lett. **89**, 137006 (2002).

[2] Tadataka Watanabe, Minoru Nohara, Tetsuo Hanaguri, and Hidenori Takagi, Phys. Rev. Lett. **92**, 147002 (2004).

[3] P Raychaudhuri, D Jaiswal-Nagar, Goutam Sheet, S Ramakrishnan, and H Takeya, Phys. Rev. Lett. **93**, 156802 (2004).

[4] T Yokoya, T Kiss, T Watanabe, S Shin, M Nohara, H Takagi, and T Oguchi, Phys. Rev. Lett. **85**, 4952 (2000).

[5] K Izawa, A Shibata, Yuji Matsuda, Y Kato, H Takeya, K Hirata, C J van der Beek, and M Konczykowski, Phys. Rev. Lett. **86**, 1327 (2001).

[6] Tuson Park, M B Salamon, Eun Mi Choi, Heon Jung Kim, and Sung-Ik Lee, Phys. Rev. Lett. **90**, 177001 (2003); Tuson Park et al., Phys. Rev. Lett. **92**, 237002 (2004).

[7] Etienne Boaknin, R W Hill, Cyril Proust, C Lupien, Louis Taillefer and P C Canfield, Phys. Rev. Lett. **87**, 237001 (2001).

[8] P Martínez-Samper, H Suderow, S Vieira, J P Brison, N Luchier, P Lejay, and P C Canfield, Phys. Rev. B **67**, 014526 (2003).

[9] K Maki, P Thalmeier, and H Won, Phys. Rev. B **65**, 140502(R) (2002).

[10] L F Mattheiss, Phys. Rev. B **49**, R13279 (1994).

[11] J I Lee, T S Zhao, I G Kim, B I Min, and S J Youn, Phys. Rev. B **50**, 4030 (1994).

[12] H Kim, C D Hwang, and J Ihm, Phys. Rev. B **52**, 4592 (1995).

[13] T Terashima, H Takeya, S Uji, K Kadowaki, and H Aoki, Solid State Commun. **96**, 459 (1995).

[14] R S Gonnelli, D Daghero, G A Ummarino, V A Stepanov, J Jun, S M Kazakov, and J Karpinski, Phys. Rev. Lett. **89**, 247004 (2002).

[15] S. V Shulga, S -L Drechsler, G Fuchs, K -H Müller, K Winzer, M Heinecke, and K Krug, Phys. Rev. Lett. **80**, 1730 (1998).

[16] T Terashima, C Haworth, H Takeya, S Uji, H Aoki and K Kadowaki, Phys. Rev. B **56**, 5120 (1997).

[17] The quality of the crystals was confirmed from the observation de Haas–van Alphen (dHvA) oscillations, T. Terashima (private communications).

[18] The contact diameter estimated from the contact resistance was ~30-40Å, whereas the mean free path estimated from the residual resistivity ($\rho_0$~4$\mu\Omega$-cm) was ~110Å. For details, see ref. 3

[19] G E Blonder, M Tinkham, and T M Klapwijk, Phys. Rev. B **25**, 4515 (1982).

[20] A Plecenik, M Grajcar, S Benacka, P Seidel and A Pfuch, Phys. Rev. B **49**, 10016 (1994).

[21] Within our experimental resolution, we cannot resolve between impurity broadening of the spectrum and a broadening arising from a distribution of gap values, since they have very similar effect on the superconducting density of states. For further details see ref. 3.

[22] The statistical variation of the superconducting energy gaps was measured for over 20 contacts in both current directions. For different point contacts $\Delta_{I\|c}$ ranges from 1.7-2.2meV and $\Delta_{I\|a}$ ranges from 0.25-0.37meV. Similarly the statistical variation in ($\Delta/\Gamma$) ranges from 0.1 to 0.3 for I∥c and 0.4 to 0.6 for I∥a. These distributions are largely from our inability to control the current direction precisely.

[23] I I Mazin, Phys. Rev. Lett. **83**, 1427 (1999).

[24] Uncertainty on the wave-vector ***k*** would also arise from the Heisenberg Uncertainty Relation, $\Delta x \cdot \Delta p \sim \hbar/2$. For a typical point contact diameter, this uncertainty is however less than 1% of $k_F$, the momentum wave-vector at Fermi level. For further details see EPAPS.

[25] In principle one should also take the band structure of the normal metal tip into consideration. However, for silver tip (used in this work) the FS nearly spherical and

isotropic. It is therefore sufficient to only take the aniostropy of the FS of the superconductor into consideration.

[26] For the temperature dependence of $H_{c2}$ see EPAPS.

[27] The behavior of $\Delta_{I\|a}$ and $\Delta_{I\|c}$ have been verified for more than 6 contacts.

[28] R S Gonnelli, D Daghero, A Calzolari, G A Ummarino, and Valeria Dellarocca, V A Stepanov, J Jun, S M Kazakov, and J Karpinski, Phys. Rev. B **69**, 100504(R) (2004).

[29] It has been argued that a two band model provides a minimal qualitative description of a multiband superconductor; *see* ref.15.

[30] A E Koshelev and A A Golubov, Phys. Rev. Lett. **90**, 177002 (2003); this calculation assumes that the two bands are in the dirty limit. However, qualitatively the same conclusions are reached when the two bands are in the clean limit (*see,* ref. 31).

[31] N Nakai, M Ichioka, and K Machida, J. Phys. Soc. Jpn. **71**, 23 (2002).

[32] For the temperature variation of the spectra for I‖a and I‖c, see, EPAPS.

[33] Qingshan Yuan and Peter Thalmeier, Phys. Rev. B **68**, 174501 (2003).

[34] H Suhl, B T Matthias, and L R Walker, Phys. Rev. Lett. **3**, 552 (1959); E J Nicol, J P Carbotte, cond-mat/0409335 (unpublished).

[35] S B Dugdale, M A Alam, I Wilkinson, R J Hughes, I R Fisher, P C Canfield, T Jarlborg, and G Santi, Phys. Rev. Lett. **83**, 4824 (1999).

[36] S L Drechsler et al., Physica C **364-365**, 31 (2001); S L Drechsler et al., Physica C **317-318**, 117 (1999).

[37] This is at variance with ref. 36 where the square FS sheet centered around P was conjectured to be gapless. We cannot explain the observation of the large gap along $k_z$ if the square sheet is gapless.